# The Sub-Geometric Phases in Density Matrix

Zheng-Chuan Wang

Department of Physics & CAS Center for Excellence in Topological Quantum Computation, The University of Chinese Academy of Sciences, P. O. Box 4588, Beijing 100049, China.

**Abstract**

In this letter, the generalization of geometric phase in density matrix is presented, we show that the extended sub-geometric phase have unified expression whatever in adiabatic or nonadiabatic procedure, the relations between them and the usual Berry phase or Aharonov-Anandan phase are established. We also demonstrated the influence of sub-geometric phases on the physical observables. Finally, our treatment is naturally used to investigate the geometric phase in mixed state.

**PACS**: 03.65.Vf, 03.65.Ge, 05.30.-d

Geometric phase[1,2], since it's discovery in 1984, had been widely explored whatever in theories or experiments[3]. Nowadays, it has played an important role in many fields[4], such as condensed matter physics, atomic and molecular physics, optics, quantum field as well as topological physics et al.. Generally, geometric phase was investigated by the Schrödinger equation, it can be studied by the path integral formalism[5] and relativistic Dirac equation[6], too. In this letter, our method is dissimilar to these works, we focus attention on the density matrix of the system, rather than on the wave function. As we know, the usual geometric phase, both the Berry phase[1] in the adiabatic evolution of wave function $|\psi(t)\rangle = e^{i\int_{R(0)}^{R(t)} i\langle\psi_n(R)|\nabla_R|\psi_n(R)\rangle dR} e^{-\frac{i}{\hbar}\int_0^t E_n(R(t'))dt'} |\psi_n(R(t))\rangle$ with instantaneous Eigen state $|\psi_n(R(t))\rangle$ and Eigen energy $E_n(R(t))$ and the Aharonov-Anandan (AA) phase[7] in the nonadiabatic evolution of wave function $|\psi(t)\rangle = e^{i\int_0^t i\langle\tilde{\psi}(t')|\frac{\partial}{\partial t'}|\tilde{\psi}(t')\rangle dt'} e^{-\frac{i}{\hbar}\int_0^t \langle\psi(t')|H(t')|\psi(t')\rangle dt'} |\tilde{\psi}(t)\rangle$ with Hamiltonian $H(t)$, have no contribution to their density matrixes, because they are cancelled by their conjugate, that motivates us to find the geometric phase which can appear in the density matrix, and is capable for us to address the effects of geometric phase on the physical observables.

It is known, one usually observes the geometric phase in the wave function by the interference experiments. Does the geometric phase produce other observable effects? Several authors had paved the way toward this goal. In 1983, Thouless et al. had already shown the influence of geometric phase on the adiabatic particle transport process[8], they also elaborated the quantized Hall conductance by means of geometric phase[9]. Based on the path integral formalism, Huratsuji et al. manifested that the geometric phase may affect the quantization rule[5], hence the quantized energy. In order to illuminate the influence of geometric phase on the acoustoelectric current induced by surface acoustoelectric wave, we developed a position dependent geometric phase, and exploited it to study this quantized acoustoelectric current[10]. Further, can we expect this geometric phase have effect on other physical observables? To our knowledge, it is an open question and worthwhile to explore. As pointed above, the usual geometric phase doesn't appear in the density matrix, but the physical observables are always calculated through the statistical average by density matrix, which tell us must find the geometric phase that can exist in density matrix.

Initially, we study the adiabatic evolution of a system $H(R(t))$ with parameter $R(t)$, whose Eigen equation is $H(R(t))|\psi_n(R(t))>= E_n(R(t))|\psi_n(R(t))>$. In order to obtain the Berry phase, we should calculate the instantaneous Eigen state $|\psi_n(R(t))>$,

however, it is seem not trivial, because the instantaneous Eigen equation is not easy to solve except some special cases, we will discuss it below.

If we denote the Hamiltonian at initial time as $H(R(0))$ with it's Eigen equation $H(R(0))|\psi_n^0> = E_n^0|\psi_n^0>$, the time-dependent Hamiltonian can be divided as

$$H(R(t)) = H(R(0)) + \Delta H(R(t)), \qquad (1)$$

where $\Delta H(R(t))$ describes the difference between Hamiltonian $H(R(t))$ and it's initial value $H(R(0))$, which should be small only if the system varies adiabatically. Then we perform the expansion of the instantaneous Eigen function by the complete basis $\{|\psi_n^0>\}$ of $H(R(0))$ as

$$|\psi_n(R(t))> = \sum_k c_{nk}(R(t)) e^{-i\omega_k t} |\psi_k^0>, \qquad (2)$$

where $\omega_k = \frac{E_k}{\hbar}$, and the coefficient $c_{nk}(R(t))$ satisfies

$$i\hbar \dot{c}_{nk}(R(t)) = \sum_l e^{-i\omega_{kl} t} <\psi_k^0|\Delta H(R(t))|\psi_l^0> c_{nl} \qquad (3)$$

with $\omega_{kl} = \frac{E_k}{\hbar} - \frac{E_l}{\hbar}$. In the case of adiabatic evolution, $\Delta H(R(t))$ may be regarded as perturbation, so $c_{nk}$ may take the form of the first order approximation as

$$c_{nk}(R(t)) = \frac{1}{i\hbar} \int_0^t e^{i\omega_{nk} t'} <\psi_k^0|\Delta H(R(t'))|\psi_k^0> dt'.$$

Remembering the expansion (2), if we introduce the notation $|\varphi_{nk}(R(t))>$ represents the basis $c_{nk}(R(t))|\psi_k^0>$, and let

$e^{if(t)}e^{-i\omega_k t}|\varphi_{nk}(R(t))>$ satisfies the Schrödinger equation, $f(t)$ must be chosen as

$$f(t) = i\int_{R(0)}^{R(t)} \frac{<\varphi_{nk}(R)|\nabla_R|\varphi_{nk}(R)> dR}{<\varphi_{nk}(R)|\varphi_{nk}(R)>} - \frac{1}{\hbar}\int_0^t \frac{<\varphi_{nk}(R(t'))|\Delta H(R(t'))|\varphi_{nk}(R(t'))> dt'}{<\varphi_{nk}(R(t'))|\varphi_{nk}(R(t'))>}, \quad (4)$$

then $\{e^{if(t)}e^{-i\omega_k t}|\varphi_{nk}(R(t))>\}$ constitute a complete basis, the expansion of the instantaneous Eigen function can also be carried out by them as

$$|\psi_n(R(t))> = \sum_k a_{nk} e^{i\int_{R(0)}^{R(t)} \frac{i<\varphi_{nk}(R)|\nabla_R|\varphi_{nk}(R)> dR}{<\varphi_{nk}(R)|\varphi_{nk}(R)>}} e^{-i(\omega_k t + \frac{1}{\hbar}\int_0^t \frac{<\varphi_{nk}(R(t'))|\Delta H(R(t'))|\varphi_{nk}(R(t'))> dt'}{<\varphi_{nk}(R(t'))|\varphi_{nk}(R(t'))>})} c_{nk}(R(t))|\psi_k^0> \quad (5)$$

where the coefficient $a_{nk}$ can be absorbed into $|\varphi_{nk}(R(t))>$, so we will not write them out in the following. The phase $\int_{R(0)}^{R(t)} \frac{i<\varphi_{nk}(R)|\nabla_R|\varphi_{nk}(R)> dR}{<\varphi_{nk}(R)|\varphi_{nk}(R)>}$ in the above expression is just the geometric phase associated with the basis $|\varphi_{nk}(R(t))> = c_{nk}(R(t))|\psi_k^0>$, we call it sub-geometric phase, while the phase $\omega_k t + \frac{1}{\hbar}\int_0^t \frac{<\varphi_{nk}(R(t'))|\Delta H(R(t'))|\varphi_{nk}(R(t'))> dt'}{<\varphi_{nk}(R(t'))|\varphi_{nk}(R(t'))>}$ is the dynamical phase. Following Expression (5), we can write it's corresponding density matrix as

$$\rho_n = \sum_{kl} e^{i(\int_{R(0)}^{R(t)} \frac{i<\varphi_{nk}(R)|\nabla_R|\varphi_{nk}(R)> dR}{<\varphi_{nk}(R)|\varphi_{nk}(R)>} - \int_{R(0)}^{R(t)} \frac{i<\varphi_{nl}(R)|\nabla_R|\varphi_{nl}(R)> dR}{<\varphi_{nl}(R)|\varphi_{nl}(R)>})} e^{-i(\omega_{kl} t + \frac{1}{\hbar}\int_0^t \frac{<\varphi_{nk}(R(t'))|\Delta H(R(t'))|\varphi_{nk}(R(t'))> dt'}{<\varphi_{nk}(R(t'))|\varphi_{nk}(R(t'))>} - \frac{1}{\hbar}\int_0^t \frac{<\varphi_{nl}(R(t'))|\Delta H(R(t'))|\varphi_{nl}(R(t'))> dt'}{<\varphi_{nl}(R(t'))|\varphi_{nl}(R(t'))>})} c_{nl}^*(R(t))c_{nk}(R(t))|\psi_k^0><\psi_l^0| \quad . \quad (6)$$

The phase $\int_{R(0)}^{R(t)} \frac{i<\varphi_{nk}(R)|\nabla_R|\varphi_{nk}(R)> dR}{<\varphi_{nk}(R)|\varphi_{nk}(R)>} - \int_{R(0)}^{R(t)} \frac{i<\varphi_{nl}(R)|\nabla_R|\varphi_{nl}(R)> dR}{<\varphi_{nl}(R)|\varphi_{nl}(R)>}$ is just the relative

sub-geometric phase between states $|\psi_k^0>$ and $|\psi_l^0>$ in density matrix.

By use of Expansion (2), the sub-geometric phase can be connected with the usual Berry phase which could be expressed as $i\int_{R(0)}^{R(t)} <\psi_n(R)|\nabla_R|\psi_n(R)>dR = i\int_{R(0)}^{R(t)} \sum_k c_{nk}^*(R(t))\frac{\partial}{\partial R}c_{nk}(R(t))dR$, while the sub-geometric phase is $i\int_{R(0)}^{R(t)} \frac{c_{nk}^*(R(t))\frac{\partial}{\partial R}c_{nk}(R(t))}{c_{nk}^*(R(t))c_{nk}(R(t))}dR$, we can find that Berry phase can be regarded as the summation of sub-geometric phases with certain probability,

$$i\int_{R(0)}^{R(t)} <\psi_n(R)|\nabla_R|\psi_n(R)>dR = i\int_{R(0)}^{R(t)} \sum_k |c_{nk}|^2 \frac{c_{nk}^*(R(t))\frac{\partial}{\partial R}c_{nk}(R(t))}{c_{nk}^*(R(t))c_{nk}(R(t))}, \quad (7)$$

which relates the Berry phase and the sub-geometric phase proposed by us.

In analogy with the adiabatic evolution, the above sub-geometric phase is suitable for the description of nonadiabatic case, in which the Hamiltonian $H(t)$ can also be written as $H(t) = H(0) + \Delta H(t)$, and it's wave function may still be expanded similar to Eq.(5) as

$$|\psi(t)> = \sum_k e^{i\int_0^t \frac{i<\varphi_k(t')|\frac{\partial}{\partial t'}|\varphi_k(t')>dt'}{<\varphi_k(t')|\varphi_k(t')>}} e^{-i(\omega_k t + \frac{1}{\hbar}\int_0^t \frac{<\varphi_k(t')|\Delta H(R(t'))|\varphi_k(t')>dt'}{<\varphi_k(t')|\varphi_k(t')>})} c_k(t)|\psi_k^0> \quad (8)$$

where $\int_0^t \frac{i<\varphi_k(t')|\frac{\partial}{\partial t'}|\varphi_k(t')>dt'}{<\varphi_k(t')|\varphi_k(t')>}$ is the sub-geometric phase, we can see that it has the same form as the adiabatic case. It should be emphasized that $\Delta H(t)$ is not small here in the nonadiabatic procedure at all, so the coefficients $\{c_k(t)\}$ in the above expression can not be obtained by the perturbative theory as adiabatic case, but they still satisfy Eq. (3), we can solve them from these coupled differential equations.

Beyond this, the sub-geometric phase in the density matrix is also analogous to the adiabatic case (6), in which the relative geometric phase $\int_0^t \frac{i<\varphi_k(t')|\frac{\partial}{\partial t'}|\varphi_k(t')>dt'}{<\varphi_k(t')|\varphi_k(t')>} - \int_0^t \frac{i<\varphi_l(t')|\frac{\partial}{\partial t'}|\varphi_l(t')>dt'}{<\varphi_l(t')|\varphi_l(t')>}$ indicates the coherence between states $|\psi_k^0>$ and $|\psi_l^0>$. Therefore, the sub-geometric phase reported in this letter have unified expressions whatever in adaibatic or nonadiabatic procedure.

Generally, the nonadiabatic AA phase $\int_0^t <\hat{\psi}(t')|i\frac{\partial}{\partial t'}|\hat{\psi}(t')>dt'$ is not easy to calculate owing to the unknown function $|\hat{\psi}(t)>$, but it can be explicitly expressed by the coefficient $c_k(t)$ in our expansion (2), where the total phase is $e^{i\phi} = <\psi(0)|\psi(\tau)>$, yielding $\phi = -i \ln \sum_n c_n^*(0) c_n(\tau) e^{-i\frac{E_n \tau}{\hbar}}$, while the dynamical phase is

$$\alpha(\tau) = -\frac{1}{\hbar} \int_0^\tau \sum_n |c_n(t)|^2 E_n dt - \int_0^\tau dt \sum_{nm} c_n^*(t) c_m(t) e^{-i\frac{(E_m - E_n)}{\hbar}} < \psi_n^0 | \frac{\Delta H(t)}{\hbar} | \psi_m^0 > \qquad (9)$$

then the AA phase $\beta$ can be finally obtained as $\beta(\tau) = \phi - \alpha(\tau)$. Therefore we provide an alternative way to calculate the AA phase.

Since the sub-geometric phase may appear in the density matrix whatever in adiabatic or nonadiabatic procedure, it will affect the physical observables through the statistical average by the density matrix, $<A> = tr(\rho \hat{A})$. As stated above, the relative sub-geometric phases in the density matrix will not vanish, they will produce observable effects on these physical quantities, which answers the question raised at the beginning.

Employing the sub-geometric phases in density matrix, we proceed to discuss the geometric phase in mixed state. For a mixed state $\rho = \sum_k p_k \rho_k$, where $\rho_k$ is the density matrix of the pure state $|\psi_k\rangle$ ($k = 1, 2...$), $p_k$ is the probability for $|\psi_k\rangle$ appearing in the mixed state, there is no fixed relative phase, including the relative geometric phase, between different pure states at all. In spite of some treatments proposed[11-15] for the geometric phase in mixed state, we remark that it is not simple to define the geometric phase in mixed state clearly. Due to the density matrix $\rho_k$ for a pure state $|\psi_k\rangle$ comprising the sub-geometric phases therein, we can naturally define the sub-geometric phases in the density matrix of mixed state by our method. Certainly, there is still no relative geometric phase between different pure states $\{|\psi_k\rangle\}$, the sub-geometric phases only exist within each pure state.

In the next, we will illustrate the sub-geometric phase by a two states system with Hamiltonian $H(t) = H_0 + H'(t)$, where

$$H_0 = \begin{pmatrix} -\Delta & 0 \\ 0 & \Delta \end{pmatrix}, \quad H'(t) = \begin{pmatrix} 0 & w(t) \\ w^*(t) & 0 \end{pmatrix} \text{ with } w(t) = |w(t)| e^{i\delta(t)}.$$

For it's instantaneous Eigen state $|\psi_-(t)\rangle = \cos\frac{\theta}{2}|\phi_1\rangle + e^{-i\delta}\sin\frac{\theta}{2}|\phi_2\rangle$, where $|\phi_i\rangle$ ($i = 1, 2$) represents the Eigen function of $H_0$ and $\tan\theta(t) = \frac{|w(t)|}{\Delta}$, the

density matrix including with sub-geometric phases is

$$\rho = \cos^2\frac{\theta}{2} |\phi_1\rangle\langle\phi_1| + e^{i(\int_0^t \frac{-\frac{1}{2}\sin\frac{\theta}{2}\cos\frac{\theta}{2}\frac{\partial}{\partial t'}\theta}{\cos^2\frac{\theta}{2}}dt' - \int_0^t \frac{\frac{1}{2}\sin\frac{\theta}{2}\cos\frac{\theta}{2}\frac{\partial}{\partial t}\theta - \sin^2\frac{\theta}{2}\frac{\partial}{\partial t'}\delta}{\sin^2\frac{\theta}{2}}dt')}$$

$$\cos\frac{\theta}{2}\sin\frac{\theta}{2}e^{i\delta} |\phi_1\rangle\langle\phi_2| + e^{i(\int_0^t \frac{\frac{1}{2}\sin\frac{\theta}{2}\cos\frac{\theta}{2}\frac{\partial}{\partial t'}\theta - \sin^2\frac{\theta}{2}\frac{\partial}{\partial t'}\delta}{\sin^2\frac{\theta}{2}}dt' - \int_0^t \frac{-\frac{1}{2}\sin\frac{\theta}{2}\cos\frac{\theta}{2}\frac{\partial}{\partial t'}\theta}{\cos^2\frac{\theta}{2}}dt')}, \quad (10)$$

$$\cos\frac{\theta}{2}\sin\frac{\theta}{2}e^{-i\delta} |\phi_2\rangle\langle\phi_1| + \sin^2\frac{\theta}{2}|\phi_2\rangle\langle\phi_2|$$

$\int_0^t \frac{-\frac{1}{2}\sin\frac{\theta}{2}\cos\frac{\theta}{2}\frac{\partial}{\partial t'}\theta}{\cos^2\frac{\theta}{2}}dt' - \int_0^t \frac{\frac{1}{2}\sin\frac{\theta}{2}\cos\frac{\theta}{2}\frac{\partial}{\partial t}\theta - \sin^2\frac{\theta}{2}\frac{\partial}{\partial t'}\theta}{\sin^2\frac{\theta}{2}}dt'$ is the relative sub-geometric phase between states $|\phi_1\rangle$ and $|\phi_2\rangle$, which is suitable for the procedure no matter what adiabatic or nonadiabatic case. While the usual Berry phase is $\beta_{Berry} = \int_0^t \sin^2\frac{\theta}{2}\frac{\partial}{\partial t'}\delta dt'$, which implies that the Berry phase is the summation of our sub-geometric phases with certain probability. The AA phase $\beta_{AA} = \phi - \alpha(\tau)$ can also be calculated by our method, where $\phi = -i\ln[\cos\frac{\theta(0)}{2}\cos\frac{\theta(\tau)}{2} + e^{i(\delta(0)-\delta(\tau))}\sin\frac{\theta(0)}{2}\sin\frac{\theta(\tau)}{2}]$ is the total phase, $\alpha(\tau) = -\int_0^\tau \sqrt{\Delta^2 + |w(t)|^2}dt$ is the dynamical phase.

This work are supported by the National Natural Science Foundation of China (Grant No. 11274378), and the Key Research Program of the Chinese Academy of Sciences (Grant No. XDPB08-3).